\documentclass{aastex}          
\usepackage{spr-astr-addons}    

\usepackage{graphicx}
\usepackage{amssymb,amsmath}
\usepackage{url}

\def\f{\frac}
\def\a{\alpha}
\def\b{\beta}
\def\e{\epsilon}
\def\d{\rm d}
\def\be{\begin{equation}}
\def\ee{\end{equation}}
\def\bea{\begin{eqnarray}}
\def\eea{\end{eqnarray}}
\def\l{\left}
\def\r{\right}

\def\no{\nonumber}

\def\s{\rm s}
\def\m{\rm m}

\begin{document}
%
\title{Tachyonic (phantom) power-law cosmology}


\author{Rachan Rangdee\altaffilmark{1}}
\email{rachanr52@email.nu.ac.th}
\and
\author{Burin Gumjudpai\altaffilmark{1,2,\dag}}
\email{buring@nu.ac.th} 

\altaffiltext{1}{The Institute for Fundamental Study ``The Tah Poe Academia Institute", Naresuan University, Phitsanulok 65000, Thailand}
\altaffiltext{2}{Thailand Center of Excellence in Physics, Ministry of Education, Bangkok 10400, Thailand}
\altaffiltext{\dag}{Corresponding author}

\begin{abstract}
Tachyonic scalar field-driven late universe with dust matter content is considered. The cosmic expansion is modeled with power-law and phantom power-law expansion at late time, i.e. $z \lesssim 0.45$. WMAP7 and its combined data are used to constraint the model.
 The forms of potential and the field solution are different for quintessence and tachyonic cases. Power-law cosmology model (driven by either quintessence or tachyonic field) predicts unmatched equation of state parameter to the observational value, hence the power-law model is excluded for both quintessence and tachyonic field. In the opposite, the phantom power-law model predicts agreeing valued of equation of state parameter with the observational data for both quintessence and tachyonic cases, i.e. $w_{\phi, 0}  = -1.49^{+11.64}_{-4.08}$ (WMAP7+BAO+$H_0$) and  $ w_{\phi, 0}  = -1.51^{+3.89}_{-6.72} $  (WMAP7).  The phantom-power law exponent $\b$ must be less than about -6, so that the $-2 < w_{\phi, 0} < -1$. The phantom power-law tachyonic potential is reconstructed. We found that dimensionless potential slope variable $\Gamma$  at present is about 1.5. The tachyonic potential reduced to $V= V_0\phi^{-2}$ in the limit $\Omega_{\m, 0} \rightarrow 0$.
\end{abstract}

\keywords{power-law cosmology, tachyonic dark energy}

\maketitle

\section{Introduction}
\label{intro}

There have been clear evidences that the present universe is under accelerating expansion as observed in, e.g. the cosmic microwave background (CMB) \citep{Masi:2002hp, Larson:2010gs, arXiv:1001.4538}, large-scale structure surveys \citep{Scranton:2003in, Tegmark:2004a}, supernovae type Ia  (SNIa) \citep{Perlmutter:1997zf, Perlmutter:1999a, Riess:1998cb, Riess:1999ar, Goldhaber:2001a, Tonry:2003a, Riess:2007a, RiessGold2004, Astier:2005qq, Amanullah2010} and X-ray luminosity from galaxy clusters \citep{Allen:2004cd, Rapetti:2005a}. One prime explanation is that the acceleration is an effect of a scalar field evolving under its potential to acquire negative pressure with $p < -\rho c^2/3$ giving repulsive gravity. Form of energy with this negative pressure range is generally called dark energy \citep{Padmanabhan:2004av, Copeland:2006a, Padmanabhan:2006a}.   Scalar field is responsible for symmetry breaking mechanisms and super-fast expansion in inflationary scenario, resolving horizon and flatness problems as well as explaining the origin of structures \citep{inflation, Guth:1981a, Sato:1981a, AlbrechtSteinhardt:1982a, Linde:1982a}.  Introducing a cosmological constant into the field equation is simplest way to have dark energy \citep{cosmoconstant, Ford:1987a, Dolgov:1997a}, but it creates new problem on fine-tuning of energy density scales \citep{SahniSta2000, PeeblesRatra:2003}.  For the cosmological constant to be viable, idea of varying cosmological constant needs to be installed \citep{varcc, ShapiroSola:2009}. If dark energy is the scalar field, the field could have non-canonical kinetic part such as tachyon which is classified in a type of k-essence models \citep{ArmendarizPicon:2000dh, ArmendarizPicon:2000ah}. The tachyon field is a negative mass mode of an unstable non-BPS D3-brane in string theory \citep{242, Sen:2002a, Sen:2002b} or  a massive scalar field on anti-D3 brane \citep{Garousi:2004uf}. It was found that the tachyonic field potential must not be too steep, i.e. less steep than $V(\phi) \propto \phi^{-2}$ in order to account for the late acceleration \citep{Padmanabhan:2002cp, Bagla:2002yn, Kutasov:2003er, Abramo:2003cp, Aguirregabiria:2004xd, Copeland:2004hq}.

In this work, we considered dark energy in form of tachyonic scalar field in power-law cosmology of which the scale factor scaled as $a \propto t^{\a}$ with $0 \leq \a  \leq \infty $, corresponding to  acceleration if $\a > 1$ and in addition we also consider phantom power-law cosmology with $a \propto (t_{\s} - t)^{\b}$. In cosmic history, there were epoch when radiation or dust is dominant component in the universe for which the scale factor evolves as power-law $a \propto t^{1/2}$ and $a \propto t^{2/3}$. A universe with mixed combination of different cosmic ingredients can be modeled using power-law expansion with some approximately constant $\alpha$ during a brief period of cosmic time. Adjustability of expansion rate is characterized by only one parameter, $\alpha$ which is used widely in astrophysical observations. There are also other situations that one can obtain the power-law solution. These are such as
non-minimally coupled scalar-tensor theory in which the scalar field couples to the curvature contributing to energy density that cancels out the vacuum energy  \citep{Dolgov1982, Ford:1987a, FujiiNishioka:1990, Dolgov:1997a} and simple inflationary model in which the power-law cosmology can avoid flatness and horizon problems and can give simple spectrum \citep{Lucchin}. The power-law has proved to be a very good phenomenological description of the cosmic evolution, since it can describe radiation epoch, dark matter epoch, and dark energy epoch according to value of the exponent \citep{Kolb1989, Peebles:1994xt}. Previously linear-coasting cosmology, $ \a  \approx 1$ was analyzed \citep{Lohiya:1996wr, Sethi:1999sq, Dev:2000du, Dev:2002sz} with motivation from SU(2) instanton cosmology \citep{Allen:1998vx}, higher order (Weyl)  gravity \citep{Manheim}, or from scalar-tensor theories \citep{Lohiya:1998tg}.  However the universe expanding with $\alpha = 1$ \citep{Melia:2011fj} was not able to agree with observational constraint from Type Ia supernovae, Hubble rate data from cosmic chronometers and BAO \citep{Bilicki:2012ub} which indicates  that $H'(z)$=const and $q(z) = 0$ are not favored by the observations.

For a specific gravity or dark energy model, power-law cosmology is considered in $f(T)$ and $f(G)$ gravities \citep{SetareAstrophysicsSpace, SetareDarabi:2012} and in the case of which there is coupling between cosmic fluids \citep{arXiv:0803.1086}.
The power-law cosmology were also studied in context of scalar field cosmology \citep{GumjudpaiPowerlaw1, Gumjudpai:2012hs}, phantom scalar field cosmology \citep{arXiv:1008.2182}. There is also slightly different form of the power-law function which $\a$ can evolved with time so that it can parameterize cosmological observables \citep{astro-ph/0405368}.

For the power-law to be valid throughout the cosmic evolution, it is not possible with constant exponent. For example, at big bang primordial nucleosynthesis (BBN), $\a$ is allowed to have maximum value at approximately $0.55$ in order to be capable of light element abundances \citep{Kaplinghat:1998wc} \citep{Kaplinghat:2000zt}. The value is about $1/2$ at highly-radiation dominated era, about $2/3$ at highly-dust dominated era and greater than one at present.
Low value of $\a$ results in much younger cosmic age and does not give acceleration. On the other hand $\a \geq 1$ value is needed to solve age problem in the CDM model \citep{Kolb1989} without flatness and horizon problems.
In universe dominated with cold dark matter and dark energy, considering that the power-law expansion happens long after matter-radiation equality era, $z \ll 3196$ (value from \citep{Larson:2010gs}), the BBN constraint can be relaxed and large $\a$ can be allowed. We consider power-law cosmology with a brief period of recent cosmic era when dark energy began to dominate, i.e. from $z  \lesssim 0.45$  to present  using results from WMAP7 \citep{Larson:2010gs} and WMAP7+BAO+$H_0$ combined datasets \citep{arXiv:1001.4538}. There are tachyonic scalar field evolving under potential $V(\phi)$ and dust barotropic fluid (cold dark matter and baryonic matter) as two major ingredients.
We aim to test whether the power-law cosmology is still valid in the scenario of tachyonic scalar field by looking at value of the equation of state predicted by the power-law tachyonic cosmology and that of varying dark energy equation of state direct-observational result. The WMAP7 and WMAP7+BAO+$H_0$ data used here are presented in Table \ref{datatable}.
We also consider when the field is phantom, i.e. having negative kinetic term with phantom power-law expansion \citep{Caldwell:1999ew, Caldwell:2003vq}, $a \propto (t_{\s} - t)^{\b},\; \b < 0$ from $z \lesssim 0.45$ till present.
We determine tachyonic field equation of state parameter, $w_{\phi}$  and we perform parametric plot versus exponent $\b$. We then analyze  the result and conclude this work.

\section{Background cosmology and observational data}
\label{sec:1}
We consider standard FLRW universe containing dust matter (cold dark matter and baryonic matter) with tachyonic field with  Lagrangian,
\be
\mathcal{L}_{\rm tachyon} \,=\, -V(\phi) \sqrt{1 - \partial_{\mu} \phi  \partial^{\mu} \phi }
 \ee
evolving under the background Friedmann equation,
\be
H^2  =  \frac{8 \pi G}{3} \l( \rho_{\phi}  + \rho_{\rm m} \r)    -  \frac{k c^2}{a^2}
\ee
and acceleration rate,
\be\label{hdot}
\dot{H}  =
\f{\ddot{a}}{a}  - H^2
=  - \f{4 \pi G}{c^2}   \l( \rho_{\phi}c^2 + p_{\phi}  + \rho_{\rm m} c^2  +  p_{\rm m} \r)   + \frac{k c^2}{a^2}
\ee
Tachyonic field energy density and pressure are
\be\label{tphi}
\rho_{\phi} c^2 =  \frac{V(\phi)}{\sqrt{1 - \e \dot{\phi}^2}}
\ee
\be\label{tp}
 p_{\phi}  =   -  V(\phi)  \sqrt{1- \e \dot{\phi}^2}
\ee
where $\e = \pm 1$. The negative $\e$ represents the case when kinetic term of the tachyon is phantom.
The tachyonic fluid equation reads
\be\label{tachyFluid}
 \f{\e \ddot{\phi}}{1- \e \dot{\phi}^2}   + 3 H \e \dot{\phi}   +\f{V'}{V}  = 0
\ee
Using  Eq. (\ref{tphi}), (\ref{tp}) and (\ref{tachyFluid}) in the (\ref{hdot}) (dust pressure is zero), we obtain
\be\label{dotH2}
\dot{H}   =  - \f{4 \pi G}{c^2} \l(  \f{ V \e \dot{\phi}^2}{\sqrt{1- \e  \dot{\phi}^2}}   +  \rho_{\rm m} c^2     \r)  + \f{k c^2}{a^2}
\ee
Using tachyonic density (\ref{tphi}) in the Friedmann equation therefore
\be\label{Vdada}
\f{V}{\sqrt{1- \e  \dot{\phi}^2}} =  \f{3}{8 \pi G/c^2}  \l( H^2 + \f{k c^2}{a^2}   \r)   -   \rho_{\rm m}c^2
\ee
Substituting (\ref{Vdada}) into the equation (\ref{dotH2}), we obtain
\be
\dot{H}   =  - {4 \pi G} \l[    \f{3 \e \dot{\phi}^2}{8 \pi G}   \l(H^2 + \f{k c^2}{a^2} \r)  - \rho_{\rm m} \e \dot{\phi}^2 + \rho_{\rm m}   \r]   + \f{k c^2}{a^2}
\ee
which can be rewritten as
\be\label{ephidot}
\e \dot{\phi}^2 \,= \,  - \l[ \f{2 \dot{H} - (2 k c^2/a^2)   +  8\pi G \rho_{\m}}{3 H^2 +  (3 k c^2/a^2) - 8 \pi G \rho_{\m}} \r]
\ee
and hence
\be\label{oneminus}
1 - \e \dot{\phi}^2 \,= \,   \f{3 {H^2} + 2 \dot{H}   +  (k c^2/a^2) }{3 H^2 + (3 k c^2/a^2) - 8 \pi G \rho_{\m}}
\ee
We use the above expression in the equation (\ref{Vdada}), as a result we can get tachyonic potential
\bea\label{Vpot}
V \, &=&\, \l[ \f{3}{8 \pi G/c^2}  \l( H^2 + \f{k c^2}{a^2}   \r)   -   \rho_{\rm m}c^2   \r] \no \\
    & & \times\sqrt{  \f{3 {H^2} + 2 \dot{H}   +  (k c^2/a^2) }{3 H^2 + (3 k c^2/a^2) - 8 \pi G \rho_{\m}} }.
\eea
Tachyonic potential of the phantom-power law is in different form from the quintessential potential of the normal power-law cosmology.
The tachyonic equation of state, $w_{\phi} $ is, from (\ref{oneminus}),
\bea\label{wphi}
 w_{\phi}\:  & = & \:\frac{p}{\rho c^2} \: = \:- (1 - \e \dot{\phi}^2) \no \\
         \:  & = & \: - \l[  \f{3 {H^2} + 2 \dot{H}   +  (k c^2/a^2) }{3 H^2 + (3 k c^2/a^2) - 8 \pi G \rho_{\m}}\r]
\eea
This can be weighed with the dust-matter content to give effective equation of state,
$ w_{\rm eff}   =   {\rho_{\phi}  w_{\phi}}/{(\rho_{\phi}    +  \rho_{\m} )}.$
With all information above, $ w_{\rm eff}$   is expressed as
\be\label{weff}
w_{\rm eff}  \: = \:  w_{\phi}\l[ 1 - \f{8\pi G \rho_{\m}/3}{H^2 + (k c^2/a^2)} \r]
\ee
We found that the equation (\ref{wphi}) and (\ref{weff}) are the same for both quintessence scalar field  \citep{Gumjudpai:2012hs} and tachyonic field cases, albeit the $\dot{\phi}$ and $V(\phi)$ are expressed differently in both cases. That is for both quintessence and tachyonic cases, $w_{\phi}$ does not depend on the scalar field model but depends on the form of expansion function. This is also true for $w_{\rm eff, 0}$. The equation of state is also independent of the sign of $\e$ which indicates negative kinetic energy. Using power-law expansion and phantom power-law expansion into (\ref{wphi}),  one can find the present value of the equation of state, $w_{\phi, 0}$. This value is a (phantom) power-law prediction of the  $w_{\phi, 0}$. We can compare this predicted value to the  $w_{\phi, 0}$ (of varying equation of state) obtained from CMB observation.

The derived data from WMAP7+BAO+$H_0$  and WMAP7 are presented in Table \ref{datatable}.   We will set $a_0=1$ and consider flat universe $k=0$ throughout (but kept $k$ in the formulae for completeness). Dust density is defined as $\rho_{\rm m, 0} = \Omega_{\m, 0} \rho_{\rm c, 0}\,.$
Total dust fluid density at present is sum of that of all dust mater types $ \Omega_{\m, 0} = \Omega_{\mathrm{CDM}, 0} + \Omega_{\rm b, 0}$.
Present value of the critical density is $\rho_{\rm c, 0} = 3 H_0^2/8 \pi G $, and radiation density is negligible.  We take the maximum likelihood value assuming spatially flat case.
Although in deriving  $t_0$, the $\Lambda$CDM model is assumed with the CMB data, however one can estimably use $t_0$  since $w_{{\rm DE}}$  is very close to -1. In SI units, the reduced Planck mass squared is
 $M_\mathrm{P}^2 =\hbar c / 8\pi G$. In this work, we also give correction to errors on future singularity time, $t_{\s}$ (phantom power-law case) reported previously in  \citep{arXiv:1008.2182}   and improve values of  $w_{\phi, 0}$ of the phantom power-law case in  \citep{arXiv:1008.2182} and of the usual power-law case reported earlier \citep{Gumjudpai:2012hs}.

\begin{table*}
\caption{Combined WMAP7+BAO+$H_0$ and WMAP7 derived parameters (maximum likelihood) from Refs. \citep{Larson:2010gs} and \citep{arXiv:1001.4538}. Here we also calculate (with error analysis) $\Omega_{\rm m, 0} = \Omega_{\rm b,0} + \Omega_{\mathrm{CDM},0}$, critical density: $\rho_{\rm c, 0} = 3 H_0^2/8 \pi G $ and matter density: $\rho_{\rm m, 0} = \Omega_{\rm m, 0}\rho_{\rm c, 0} $. The space is flat and $a_0$ is set to unity.}
\label{datatable}
\begin{tabular}{lll}
\tableline\noalign{\smallskip}
\textbf{Parameter} & \textbf{WMAP7+BAO+$H_0$} & \textbf{WMAP7}\\
\noalign{\smallskip}\tableline\noalign{\smallskip}
$t_0$ & $13.76\pm0.11$ Gyr or $(4.34\pm0.03) \times 10^{17}$ sec & $13.79\pm0.13$  Gyr  or $(4.35\pm0.04) \times 10^{17}$ sec \\
$H_0$ & $70.4\pm1.4$ km/s/Mpc & $70.3\pm2.5$ km/s/Mpc\\
         & $(2.28 \pm 0.04)\times 10^{-18}$ sec$^{-1}$ &  $(2.28 \pm 0.08)\times 10^{-18}$ sec$^{-1}$     \\
$\Omega_{\rm b,0}$ & $0.0455\pm0.0016$ & $0.0451\pm0.0028$\\
$\Omega_{\mathrm{CDM},0}$ & $0.226\pm0.015$ & $0.226\pm0.027$\\
\noalign{\smallskip}\tableline\noalign{\smallskip}
$\Omega_{\rm m, 0}$  &  $0.271(5)\pm0.015(1)$    &  $0.271(1)\pm0.027(1)$  \\
$\rho_{\rm m, 0}$  & $(2.52(49)^{+0.18(24)}_{-0.16(64)}) \times 10^{-27}$ ${\rm kg/m^3}$ & $(2.52(12)^{+0.30(97)}_{-0.30(61)}) \times 10^{-27}$ ${\rm kg/m^3}$\\
    $\rho_{\rm c, 0}  $  & $(9.29(99)^{+0.32(92)}_{-0.32(35)})\times 10^{-27}$ ${\rm kg/m^3}$ & $(9.29(99)^{+0.66(41)}_{-0.64(12)}) \times 10^{-27}$ $ {\rm
    kg/m^3}$\\
\noalign{\smallskip}\tableline
\end{tabular}
\end{table*}

\section{Power-law cosmology}
\label{Sec_Power}
Origin of power-law cosmology comes from a solution of the Friedmann equation with flat geometry and domination of dark energy,
$ H^2 = 8 \pi G \rho_{\phi}/3$\,. For constant equation of state $w_{\phi}$, the solution is well known as (see, for example, in page 150 of \citep{coles})
\be\label{may}
a \:  =  \: a_0 \l[ 1 + \f{H(t_0)}{\a}(t -  t_0)  \r]^{\a}
\ee
where $\a = 2/[3(1+w_{\phi})]$ is constant.
For $-1/3 > w_{\phi} > -1$, the solution takes power-law form,
\begin{equation}\label{scalefactor}
 a(t) = a_0 \left( \frac{t}{t_0} \right)^{\a}\,,
\end{equation}
Note that although the function is motivated by domination of constant $w_{\phi}$ scalar field in the flat Friedmann equation which gives $ 1 < \a < \infty$, here we will consider the range $0 < \a < \infty$ (constant value of $\a$)  and we will estimably use the power-law expansion in presence of barotropic dust fluid and varying $w_{\phi}$ in a short range of redshift $z \lesssim 0.45$ to present. Later section on phantom-power law (Sec. \ref{p_art}) is based on the same estimation as well.
In the power-law cosmology, the speed is $\dot{a} = \a a/t $ and the acceleration is $\ddot{a} = \a(\a-1) a /t^2 $. The Hubble parameter is $ H(t) = \dot{a}/{a} = {\a}/{t} $ with $ \dot{H} = -\a/t^2 $.
The deceleration parameter in this scenario is
$
q\: \equiv \: - {a \ddot{a}}/{\dot{a}^2}\: =\:  ({1}/{\a}) -1,
$
that is $\a = 1/(q+1)$. As $\a \geq 0$ is required in power-law cosmology, hence $q \geq -1$ and $H_0 \geq 0$. To convert into redshift $z$, from $1+z =  a_0/a $ then $1+z =  (t_0/t)^{\a}$.
Typically astrophysical tests  for power-law cosmology indicating the value of $\a$ are performed by observing $H(z)$ data of SNIa or high-redshift objects such as distant globular clusters \citep{Dev:2008ey, Sethi2005, Kumar2011}. To indicate the value of $\a$  one can also use gravitational lensing statistics \citep{Dev:2002sz}, compact-radio source \citep{Jain2003} or using X-ray gas mass fraction measurements of galaxy clusters \citep{Zhu:2007tm} \citep{Allen2002, Allen:2003}.
Study of angular size to $z$ relation of a large sample of milliarcsecond compact radio sources in flat FLRW universe found that $\a =  1.0 \pm 0.3$ at 68 \% C.L. \citep{Jain2003}. WMAP5 dataset gives $\a =  1.01$ for closed geometry  \citep{GumjudpaiPowerlaw1}.
Some procedures of measurement give large value of $\a$ such as $\a = 2.3^{+1.4}_{-0.7} $ (X-ray mass fraction data of galaxy clusters in flat geometry) \citep{Zhu:2007tm} and $\a = 1.62^{+0.10}_{-0.09}$  (joint test using Supernova Legacy Survey (SNLS) and $H(z)$ data in flat geometry) \citep{Dev:2008ey}.
Notice that assumption of non-zero spatial curvature ($\pm 1, 0$) is assumed in these results in evaluating of  $\a$ except in the WMAP5 of which the result puts also constraint on the spatial curvature. When $\a$ is found with curvature-independent procedure (i.e. with neither SNIa nor cluster X-ray gass mass fraction) or in flat case, $\a$  is near unity. For example, $H(z)$ data gives $\a = 1.07^{+0.11}_{-0.09}$ \citep{Dev:2008ey} and  $\a = 1.11^{+0.21}_{-0.14}$  \citep{Gumjudpai:2012hs, Kumar2011}.
Short review of recent $\a$ values can be found in Ref. \citep{Gumjudpai:2012hs}.
Here $\a$ is calculated from value at present   $H_0, t_0$ as
  $\a = H_0 t_0$.
From (\ref{wphi}) and (\ref{weff}), in case of power-law cosmology driven by tachyonic field, the equation of state of dark energy is
\bea
w_{\phi}  & = &  - \,
\f{\l[ \f{3\a^2}{t^2}\, -\,  \f{2\a}{t^2}  \,+ \, \f{k c^2}{a_0^2} \l( \f{t_0}{t} \r)^{2\a} \r]}{\l[    \f{3\a^2}{t^2} +  \f{3k c^2}{a_0^2} \l( \f{t_0}{t} \r)^{2\a}
          - 8 \pi G \rho_{\m, 0}\l(\f{t_0}{t}\r)^{3\a} \r]}
\eea
and
\be
w_{\rm eff}  =  w_{\phi} \l[ 1-  \f{(8 \pi G/3) \rho_{\m, 0} (t_0/t)^{3\a}}{(\a^2/t^2) + (kc^2/a_0^2)(t_0/t)^{2\a}} \r]
\ee
At present, $t = t_0$, $w_{\rm eff, 0} = -1 + 2/(3\a)$. In Table \ref{tablePower}, values of equation of state parameters derived in the power-law cosmology (true for both tachoynic and quintessence) do not match observational data, i.e.  $w_{\phi, 0}$ and $w_{\rm eff, 0}$ found here are much greater than observational (spatially flat) WMAP derived results, for example WMAP7\footnote{flat geometry, constant  $w_{\phi, 0}$ (Sec. 4.2.5 of Ref. \citep{Larson:2010gs})}: $w_{\phi, 0} = -1.12^{+0.42}_{-0.43}$, WMAP7+BAO+$H_0$ combined\footnote{flat geometry, constant $w_{\phi, 0}$ (Sec. 5.1 of
Ref. \citep{arXiv:1001.4538})}:
$w_{\phi, 0} = -1.10^{+0.14}_{-0.14} $ (68 \% CL), WMAP7+BAO+$H_0$+SN\footnote{flat geometry, time varying dark energy EoS, $w_{\phi}(a) = w_0+w_a(1-a)$ with $w_0 = -0.93\pm0.13, w_a = -0.41^{+0.72}_{-0.71} $  (Sec. 5.3 of Ref.  \citep{arXiv:1001.4538})}:  $w_{\phi, 0} = -1.34^{+1.74}_{-0.36}$ (68 \% CL)   and WMAP7+BAO+$H_0$+SN with time delay distance information correction\footnote{flat geometry, time varying dark energy EoS, $w_{\phi}(a) = w_0+w_a(1-a)$ with $w_0 = -0.93\pm0.12, w_a = -0.38^{+0.66}_{-0.65} $   (Sec. 5.3 of Ref.  \citep{arXiv:1001.4538})}:    $w_{\phi, 0} = -1.31^{+1.67}_{-0.38}$ (68 \% CL). We conclude that the power-law expansion universe with quintessential scalar field \citep{Gumjudpai:2012hs} or tachyonic field  is neither viable.

\begin{table*}
\caption{Power-law cosmology exponent and its prediction of equation of state parameters. The value does not match the WMAP7 results.}
\label{tablePower}
\begin{tabular}{lll}
\tableline\noalign{\smallskip}
\textbf{Parameter} & \textbf{WMAP7+BAO+$H_0$} & \textbf{WMAP7}\\
\noalign{\smallskip}\tableline\noalign{\smallskip}
$\a$ & $ 0.98(95)\pm 0.01(87) $  & $ 0.99(18) \pm 0.03(60) $  \\
$w_{\phi, 0}$ (with power-law cosmology) & $ -0.44(79)^{+0.01(66)}_{-0.01(54)} $  &  $ -0.44(98)^{+0.02(97)}_{-0.02(82)} $               \\
$w_{\rm eff, 0}$ (with power-law cosmology) & $ -0.32(63) \pm {0.01(25)} $  &  $ -0.32(78) \pm {0.02(35)} $  \\
\noalign{\smallskip}\tableline
\end{tabular}
\end{table*}

\section{Phantom power-law cosmology}
\label{p_art}

\begin{figure*}
\centering
\includegraphics[width=1.5\columnwidth]{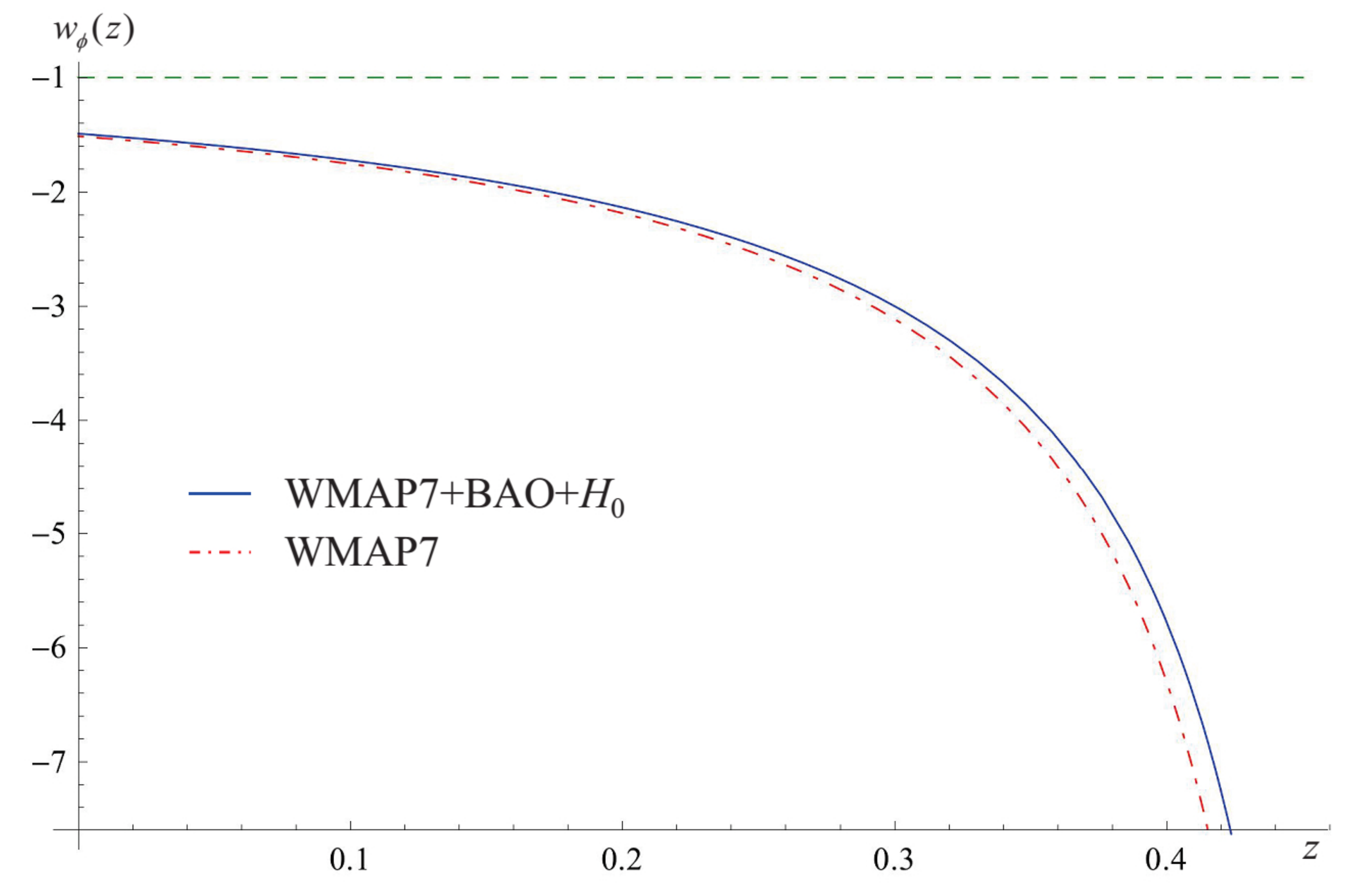}
\caption{Phantom tachyonic (and quintessence) dark energy equation of state versus $z$.  }
\label{fig:2}
\end{figure*}

In this section, we can check if phantom power-law could be a valid solution for the tachyonic-driven universe.
From Eq. (\ref{may}), with constant $w_{\phi} < -1$, the solution becomes phantom power-law,
\be
a(t) = a_0 \l(  \f{t_{\rm s} - t}{t_{\s} -t_0} \r)^{\b}
\ee
with speed,
\be
\dot{a} = - a_0 \b \f{(t_{\s} - t)^{\b -1}}{(t_{\s} - t_0)^{\b}} = - \b \f{a}{(t_{\s} - t)} \no \ee
and acceleration,
\be \ddot{a} = a_0 \b(\b-1) \f{\l(t_{\s} - t \r)^{\b-2}}{\l(t_{\s} - t_0\r)^{\b}}
 = \f{\b (\b -1) a}{(t_{\s} -t)^2} \no \ee
  where $t_{\s} \equiv t_0 + |\beta|/H(t_0)$ \citep{coles} is future big-rip singularity time \citep{Caldwell:1999ew, Caldwell:2003vq} and we use $\beta$ instead of $\a$ to distinct the two solutions. For both $\dot{a} $ and $\ddot{a} $ to be  greater than zero, i.e. both expanding and accelerating, the condition $\b < 0$ is needed. The Hubble parameter is therefore,
\be
H \,=\, \f{\dot{a}}{a} \,=\, -\f{\b}{t_{\s} - t}\;\;\;\;\;\;\; {\rm hence} \;\;\;\;\;\;\;
\dot{H}  \,=\, -\f{\b}{(t_{\s} - t)^2}
\ee
At present,  $\b =  H_0 (t_0 -t_{\s})$. The deceleration parameter  is
$ q\: \equiv \: - {a \ddot{a}}/{\dot{a}^2}\: =  \:  ({1}/{\b}) -1 $.
The dust matter density, $\rho_{\m} = \rho_{\m, 0} a_0^3 / a^3 $ is then
\be
\rho_{\m} = \rho_{\m, 0} \l(\f{t_{\s} - t_0}{t_{\s} - t}\r)^{3 \b}
\ee
Substituting these equations into (\ref{wphi}) and (\ref{weff}), we obtain,
\bea
w_{\phi} \,& = &\, - \l[    \f{\b(3\b-2)}{(t_{\s} - t)^2} \;  + \;
   \l(\f{k c^2}{a_0^2}\r) \l[\f{(t_{\s} - t_0)}{(t_{\s} - t)} \r]^{2\b}  \r] \no \\
   &   & \Bigg/ \Bigg\{   \f{3\b^2}{(t_{\s} - t)^2}    \; +    \;
    \l(\f{3k c^2}{a_0^2}\r) \l[\f{(t_{\s} - t_0)}{(t_{\s} - t)} \r]^{2\b}  \; \no \\
   &   & -  \; 8 \pi G \rho_{\m, 0} \l[\f{(t_{\s} - t_0)}{(t_{\s} - t)} \r]^{3\b}        \Bigg\}
\eea
\bea
w_{\rm eff} \,& = &\, w_{\phi}\, \Bigg[  1  - \Bigg\{\f{8 \pi G \rho_{\m, 0}}{3} \l[\f{(t_{\s} - t_0)}{(t_{\s} - t)} \r]^{3\b} \no \\
              &   & \Bigg/ \l(\f{\b^2}{(t_{\s} - t)^2}  \;+\; \l(\f{k c^2}{a_0^2}\r)\l[\f{(t_{\s} - t_0)}{(t_{\s} - t)} \r]^{2\b}\r) \Bigg\}          \Bigg] \no \\
\eea
To convert to redshift one can use $1+z = a_0/a $ therefore
 $ 1+z = \l[ (t_{\s} - t_0)/(t_{\s} - t) \r]^{\b}\;
 $ and $ t_{\s} - t = (t_{\s} - t_0)(1+z)^{-1/\b} \;$.
At present, $t = t_0$, $w_{\rm eff, 0} = -1 + 2/(3\b)$. The big-rip time $t_{\s}$, can be calculated from
\be
t_{\s} \approx  t_0 - \f{2}{3 (1+w_{\rm DE})}  \f{1}{H_0\, \sqrt{1 - \Omega_{\m, 0}}}
\ee
Here, $w_{\rm DE}$ must be less than -1 and in deriving this above expression flat geometry and constant dark energy equation of state is assumed \citep{Caldwell:1999ew, Caldwell:2003vq}. We will estimably use $t_{\s}$ from this formula. In finding error bar of $t_{\s}$, we exploit better procedure than that performed earlier in \citep{arXiv:1008.2182} by considering that the second order of error bar multiplications are too large to be neglected. We discuss this in the appendix.  Results presented in Table \ref{tablePhan} are $\b, t_{\s}$ and the equation of state.  For phantom power-law cosmology driven by tachyonic field (also true for phantom quintessence), the resulting value is $w_{\phi, 0}  = -1.49^{+11.64}_{-4.08} $  (using WMAP7+BAO+$H_0$) and  $-1.51^{+3.89}_{-6.72} $ (using WMAP7). These do not much differ from results from WMAP7+BAO +$H_0$+SN data (flat, varying dark energy EoS) which gives $w_{\phi, 0} = -1.34^{+1.74}_{-0.36}$ (68 \% CL)  and WMAP7+BAO+$H_0$+SN +time delay distance correction data (flat varying dark energy EoS) which gives $w_{\phi, 0} = -1.31^{+1.67}_{-0.38}$ (68 \% CL) \citep{arXiv:1001.4538}. Using observational data in Tables \ref{datatable} and \ref{tablePhan} we derive
\bea
w_{\phi, 0} &=& -\l[ \f{1-2/(3\b)}{1-(16.60/\b^2)} \r] ({\rm WMAP7+BAO+}H_0) \no  \\
\\
w_{\phi, 0} &=& -\l[ \f{1-2/(3\b)}{1-(11.47/\b^2)} \r] ({\rm WMAP7})
\eea
With these, we show parametric plots of the $w_{\phi, 0}$ and $\b$ in Fig. \ref{fig:1}. The values measured for $\b$ and $w_{\phi, 0}$ are the purple cross (WMAP7+BAO+$H_0$) and yellow spot (WMAP7). For $ -\infty < \b \lesssim -6$, $w_{\phi, 0}$ lies in the range $(-1, -2)$. Fig. \ref{fig:2} shows evolution of $w(z)$ in late phantom power-law universe from $0 < z < 0.45$, i.e. $t = 8.48 $ Gyr (both datasets) till present era (this is to avoid singularity in $w_{\phi}$ at $z=0.492$ (WMAP7+BAO+$H_0$) and at $z=0.484$ (WMAP7)).
 These are equivalent to the past $5.28$ Gyr ago (WMAP7+BAO+$H_0$) and  the past $5.31$ Gyr ago (WMAP7).

\begin{table*}
\caption{Phantom power-law cosmology exponent and its prediction of equation of state parameters. The equation of state lies in acceptable range of values given by WMAP7 results. Large error bar of $w_{\phi, 0}$ is an effect of large error bar in $t_{\s}$.}
\label{tablePhan}
\begin{tabular}{lll}
\tableline\noalign{\smallskip}
\textbf{Parameter} & \textbf{WMAP7+BAO+$H_0$} & \textbf{WMAP7}\\
\noalign{\smallskip}\tableline\noalign{\smallskip}
$\b$ & $ -7.81(08)^{+11.71(8)}_{-4.56(1)} $  & $ -6.50(72)^{+3.91(92)}_{-5.09(96)} $  \\
$t_{\s}$ & $122.30(0)^{+162.83(7)}_{-63.36(0)}\;$ Gyr  & $ 104.21(5)^{+54.37(3)}_{-70.79(9)}\;$ Gyr  \\
$w_{\phi, 0}$ (with phantom power-law cosmology) & $ -1.48(99)^{+11.64(46)}_{-4.08(45)} $  &  $ -1.51(26)^{+3.89(23)}_{-6.71(90)} $               \\
$w_{\rm eff, 0}$ (with phantom power-law cosmology) & $ -1.08(54)^{+0.25(60)}_{-0.11(98)} $  &  $ -1.10(24)^{+0.15(52)}_{-0.37(12)} $  \\
\noalign{\smallskip}\tableline
\end{tabular}
\end{table*}

\section{Tachyonic potential for phantom power-law cosmology}
\label{sec:phantom-pl}

\subsection{Tachyonic field dominant case}
\begin{figure*}
\centering
\includegraphics[width=1.5\columnwidth]{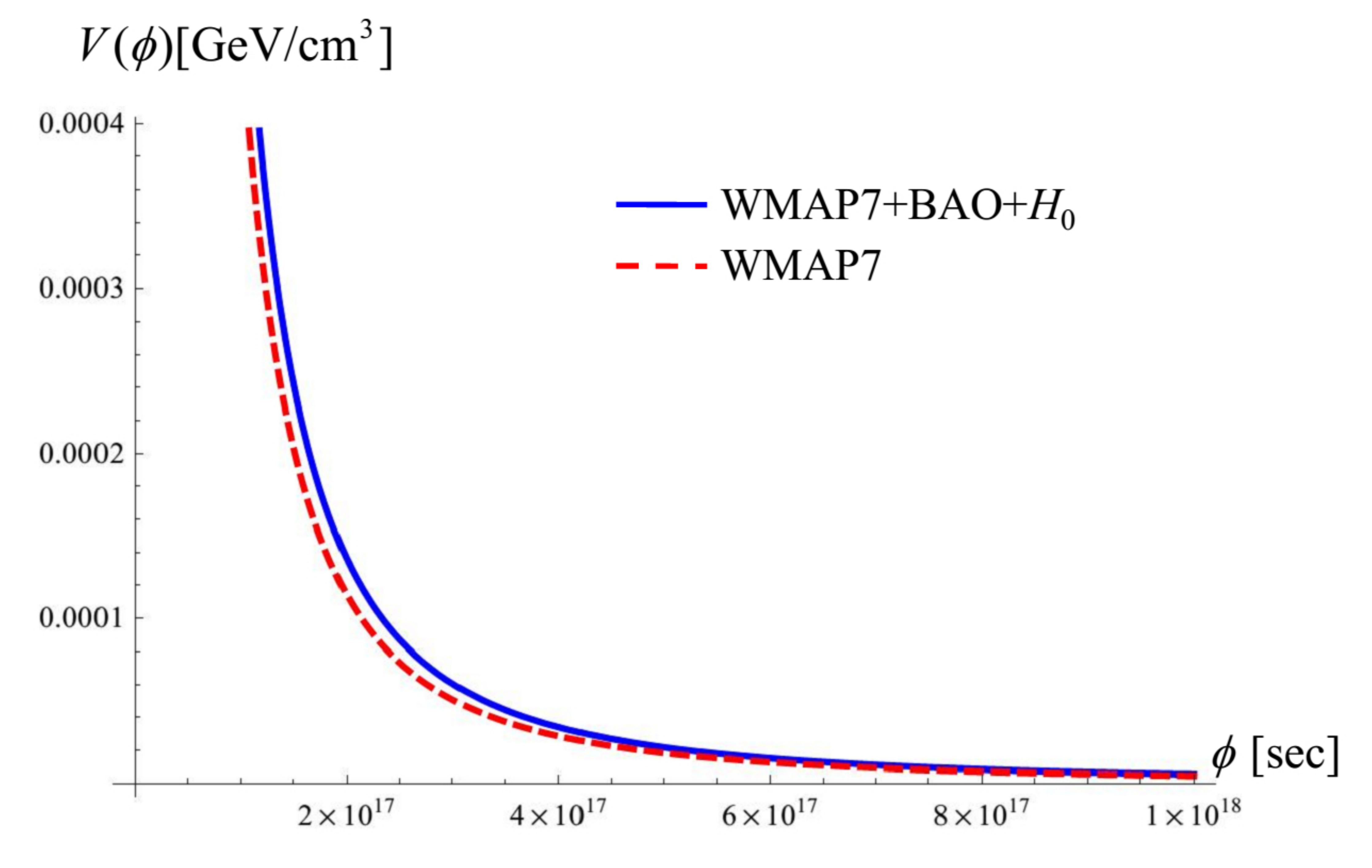}
\caption{Potential versus field  using WMAP7+BAO+$H_0$, WMAP7 for the case of tachyonic field domination ($V \propto \phi^{-2}$).}
\label{fig:31}
\end{figure*}

When the field is phantom ($\e = -1$) and is the dominant component, the equation (\ref{ephidot}) for flat space  hence
\be
 \dot{\phi}^2 \, = \,   \f{2 \dot{H}}{3 H^2} \,=\, - \f{2}{3\b}
\ee
Integrating from $t$ to $t_{\s}$, and choosing positive solution,
\be
 \phi(t) \, = \,  \sqrt{\f{2}{3|\b|}} \:\l(t_{\s} - t \r)
\ee
Since $\b < 0$ hence $-\b = |\b|$.  From (\ref{Vpot}) the tachyonic potential is
\bea
V(\phi) & = &  \f{2 c^2 |\b|}{\kappa   \phi^2   }
          \sqrt{1 + \f{2}{3|\b|}}       \label{equaDOM}
\eea
where  $ \kappa \equiv 8 \pi G $. With parameters in Table \ref{tablePhan}, the potential is plotted in Fig. \ref{fig:31} which is no surprised as it was found earlier
\citep{Padmanabhan:2002cp} regardless of the expansion is either normal power-law or phantom power-law. The steepness of the potential is tipically determined by a dimensionless variable
  \be \Gamma = \f{V'' V}{V'^2} \ee
 where $'$ denotes $\d/\d\phi$. For the potential (\ref{equaDOM}), it is found that $\Gamma = 3/2$.

\subsection{Using tachyonic field dominant solution to approximate $V(\phi)$ in mixed fluid universe}
\begin{figure*}
\centering
\includegraphics[width=1.5\columnwidth]{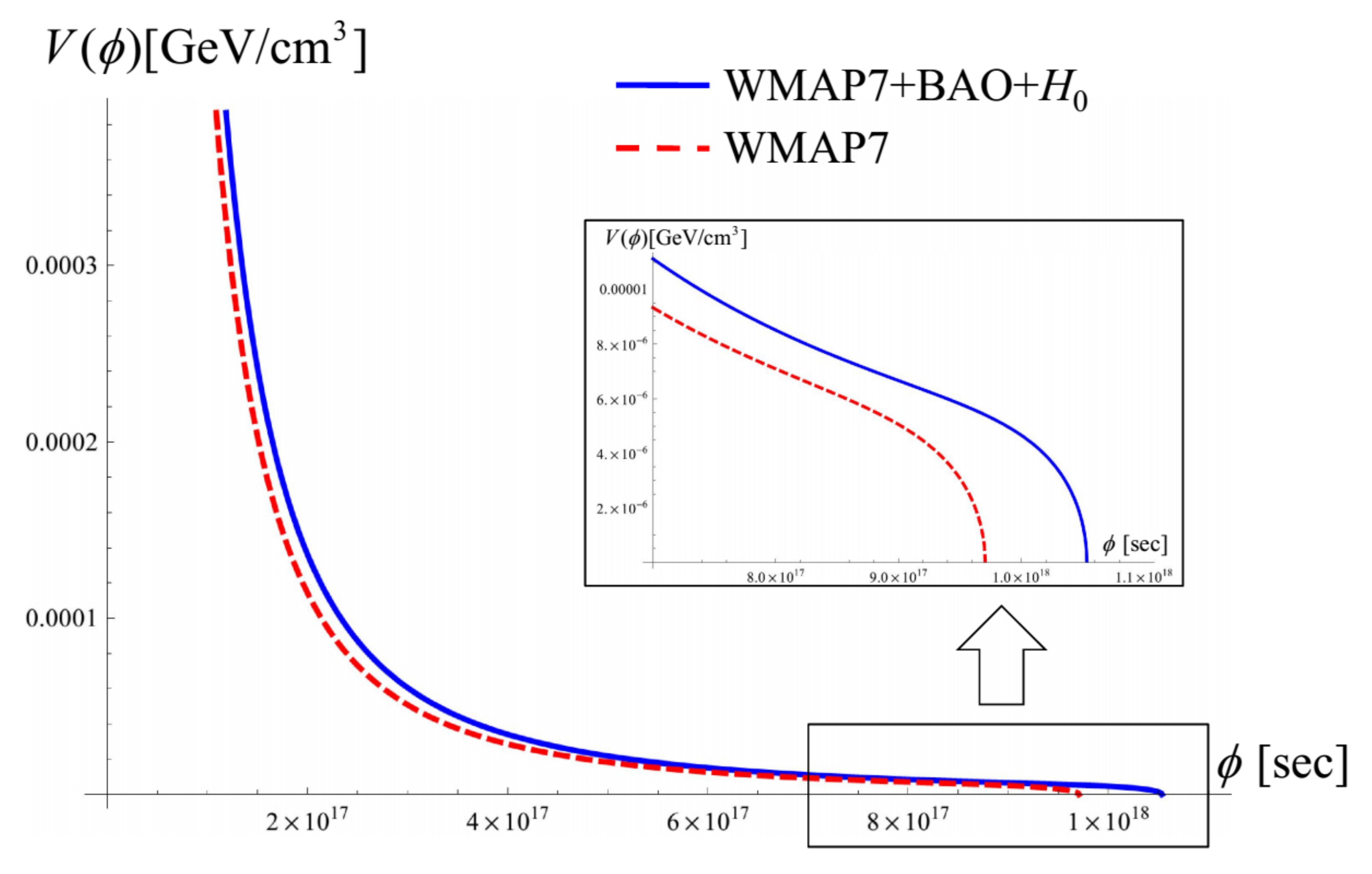}
\caption{Approximated potential versus field using WMAP7+BAO+$H_0$, WMAP7 for
the case of mixed tachyonic field with barotropic dust.}
\label{fig:3}
\end{figure*}

\begin{figure*}
\centering
\includegraphics[width=1.5\columnwidth]{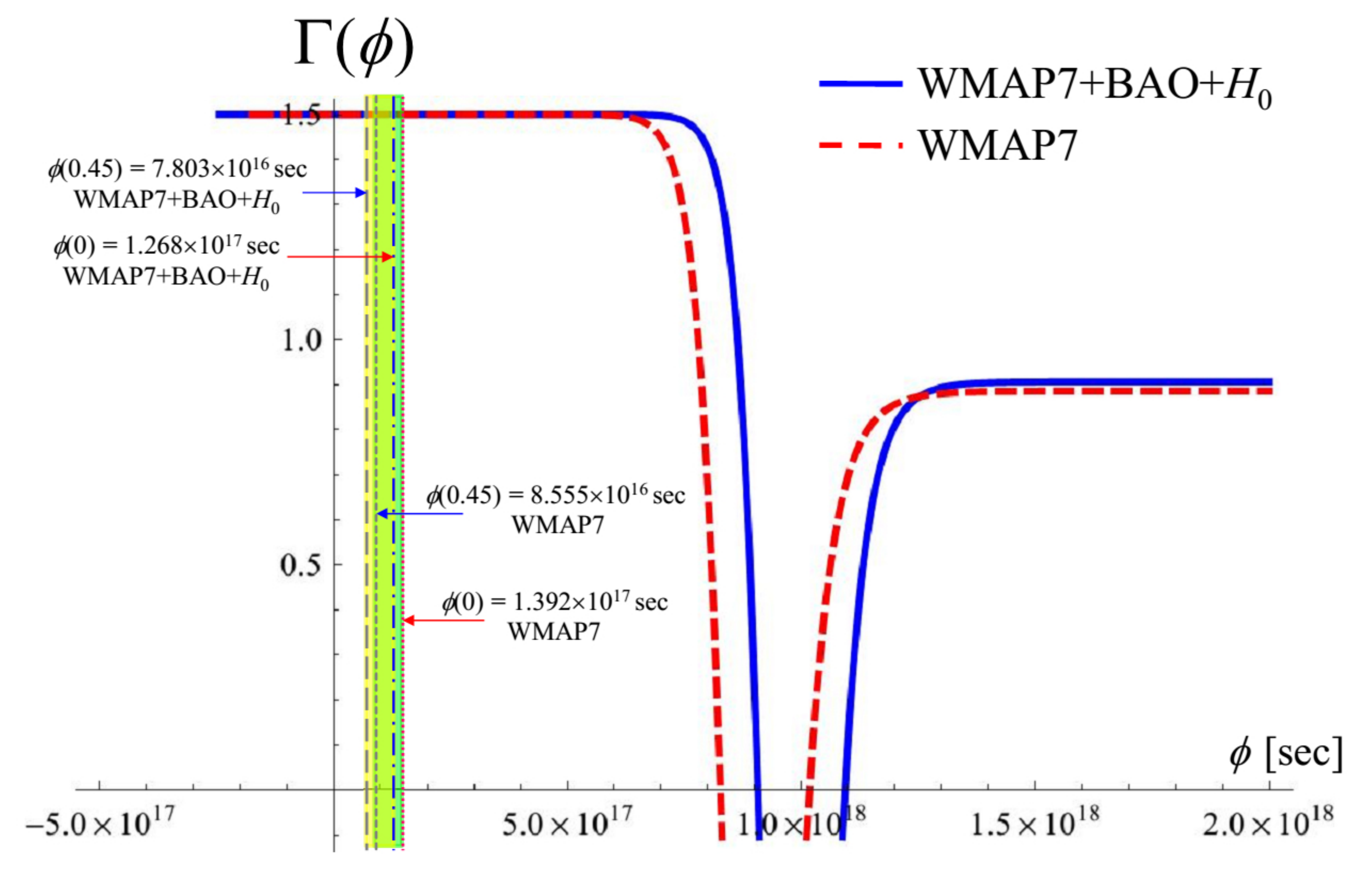}
\caption{Dimensionless variable $\Gamma$ plotted versus field using WMAP7+BAO+$H_0$ and WMAP7. The considered region for late universe $z<0.45$ lies in the bars. This is for
the case of mixed tachyonic field with barotropic dust.}
\label{fig:4}
\end{figure*}

Considering equation (\ref{ephidot}) for flat space and $\e = -1$ hence
$
 \dot{\phi}^2 \,= \,   (2 \dot{H}    +  8\pi G \rho_{\m})/(3 H^2    - 8 \pi G \rho_{\m}).
$
We approximate that the dust term is much less contributive compared to the $\dot{H}$ and $H^2$ terms therefore,
\be
 \dot{\phi}^2 \, \approx \,   \f{2 \dot{H}}{3 H^2} \,=\, - \f{2}{3\b},\;\;\;\; \phi(t) \, \approx \,  \sqrt{\f{2}{3|\b|}} \:\l(t_{\s} - t \r)
\ee
Now we will use this $\phi(t)$ solution found with tachyonic field dominant approximation to find the potential. This is not exact way of deriving the potential which has also contribution of baryonic matter density. However the approximation made here does not much alter the result and could be roughly acceptable. Let $B \equiv \sqrt{{3 |\b| }/{2}} $, hence  $t_{\s} - t  = B\phi $. Using the equation (\ref{Vpot}) we find that
\bea
V(\phi) & \approx & \l[ \f{3 c^2 \b^2}{\kappa  (B \phi )^2   }  -   \rho_{\m, 0}\, c^2  \l(    \f{t_{\s} - t_0}{ B \phi  }   \r)^{3 \b}  \r] \no \\
        &   & \times   \l[\f{1 - 2/(3\b)}{ 1  -  \rho_{\m, 0} [\kappa/(3\b^2)] \l( B \phi  \r)^{2-3\b}(t_{\s} - t_0)^{3\b} }        \r]^{1/2} \no \\
\eea
Note that the term $1 - 2/(3\b)$ is just $-w_{\rm eff, 0}$. We can rearrange the potential in form of cosmological observables $H_0, \Omega_{\m, 0}$ and $q$,

\bea
V & \approx & \frac{c^2}{\kappa}\l[ \f{2 |\b|}{\phi^2} - 3 \l(\f{3}{2 |\b| }\r)^{\f{3 |\b|}{2} }  \,\Omega_{\m, 0}\, H_0^{2+3|\b|}  {\phi}^{3 |\b|} \r] \no \\
       &&  \times \l[\f{1 + 2/(3|\b|)}{ 1  -  \l(\f{3}{2}\r)^{1+\f{3|\b|}{2}} \, \Omega_{\m, 0}\, (H_0   \phi)^{2+3|\b|}  |\b|^{-1-\f{3|\b|}{2}}}        \r]^{1/2}  \no \\  \label{Vphiphan}
\eea
where $\b = \b(q) = (1+q)^{-1}$.  This is plotted in Fig. \ref{fig:3} where the field values at present and at $z=0.45$ are
\bea
\phi|_{z=0} & = & 1.268 \times 10^{17}\;\;  {\rm sec}    \;\; {\rm and}\;\;\;  \no \\
\phi |_{z=0.45} & = & 7.803 \times 10^{16} \;\; {\rm sec} \; ({\rm WAMP7+BAO}+H_0) \no \\
\phi|_{z=0}  & = & 1.392 \times 10^{17} \;\; {\rm sec}    \;\; {\rm and} \;\;\;   \no \\
\phi |_{z=0.45} & = & 8.555 \times 10^{16}\;\;  {\rm sec} \; ({\rm WAMP7}) \no
\eea
It has been known that in order for the tachyonic potential to account for the late acceleration, it should not be steeper than the potential $V \propto \phi^{-2}$
 \citep{Padmanabhan:2002cp, Bagla:2002yn}. To check if our derived tachyonic potential could fit in this criteria, i.e. shallower than $V \propto \phi^{-2}$, we use dimensionless variable,  $\Gamma$.
For  the potential $V \propto \phi^{-2}$ in previous section,  $\Gamma = 3/2$. Hence in general the potential with $\Gamma < 3/2$ satisfies this criteria. Considering the potential (\ref{Vphiphan}) we use both derived datasets to compute its dynamical slope $\Gamma(\phi)$ which is in very complicated form. We plot this in Fig. \ref{fig:4}. We found that using our data with the field value at present, for WMAP7+BAO+$H_0$, $\Gamma(\phi(z=0)) = 1.500$ and for WMAP7,  $\Gamma(\phi(z=0)) = 1.500$. Up to three decimal digits, these values are approximately the same as that of  $V \propto \phi^{-2}$. Note that the $V \propto \phi^{-2}$ potential  is found when the universe is filled with tachyon field as single component. Indeed in the limit $\Omega_{\m, 0} \rightarrow 0$, our derived potential (\ref{Vphiphan}) becomes $V \propto \phi^{-2}$.
 The other tachyonic potentials such as $V = V_0/[{\rm cosh}(a\phi/2)]$ and $V = V_0 e^{(1/2)m^2\phi^2}$ have $\Gamma = 1 - {{\rm csch}^2(a\phi/2)}$ and $1+ (m\phi)^{-2}$ respectively. These examples are typical tachyonic potentials which also have dynamical slopes. In Fig. \ref{fig:4}, $\Gamma(\phi)$ diverges twice however, in the region we consider ($z=0.45 \rightarrow z=0$), the value of $\Gamma$ stays approximately at $1.5$.

\begin{figure*}
\centering
\includegraphics[width=1.5\columnwidth]{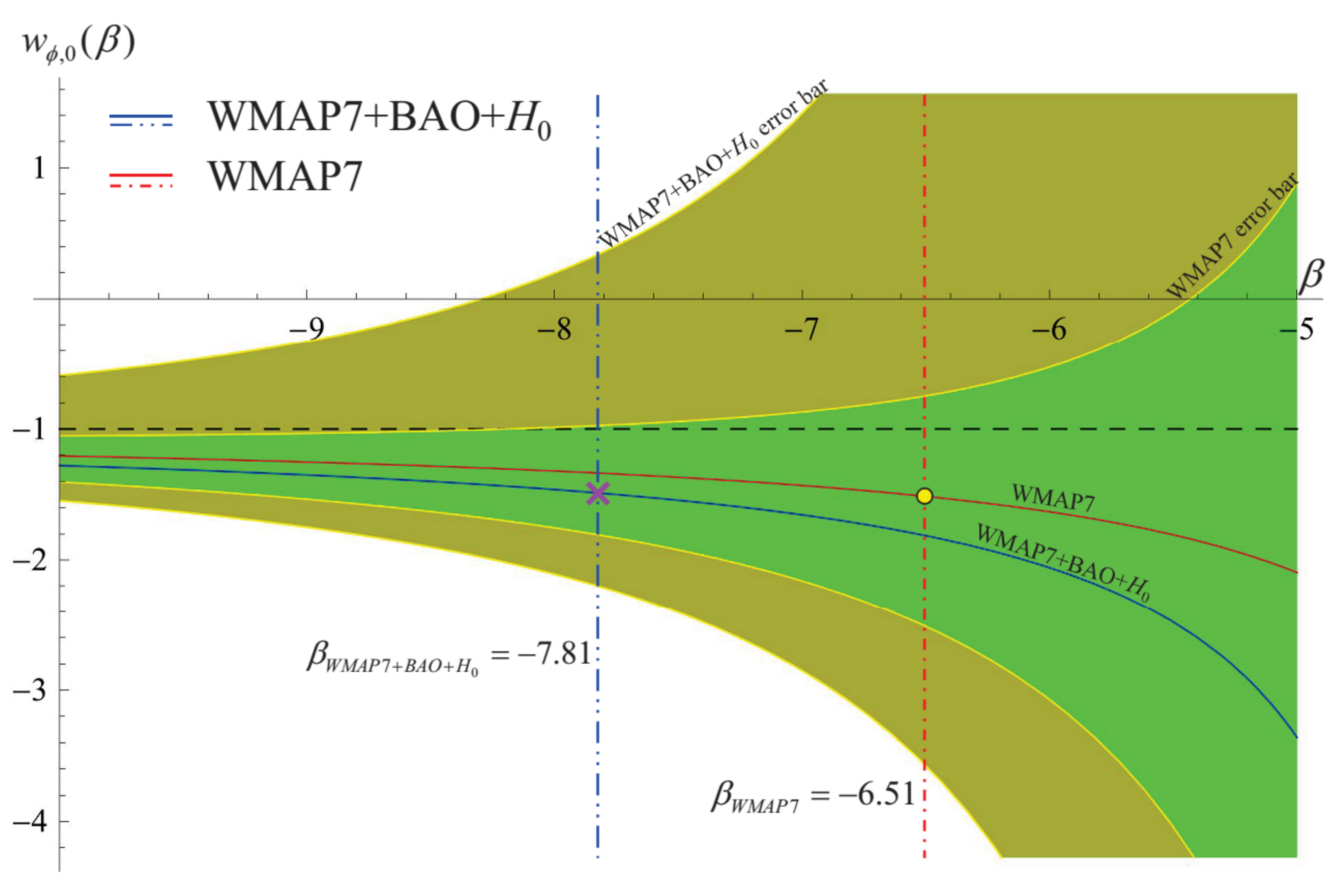}
\caption{Present value of phantom tachyonic dark energy equation of state plotted versus $\beta$.  Their error bar results from the error bar in $\beta$. This is the same for quintessence case.}
\label{fig:1}
\end{figure*}

\section{Conclusion}
In this work model of tachyonic-driven universe are investigated for normal power-law cosmology and phantom power-law cosmology. The universe is flat FLRW filled with tachyonic scalar field and dust. We consider late universe when dark energy has dominated, i.e.  $z<0.45$. WMAP7 data and its derived data when combined with BAO and $H(z)$ data are used in this study. We find exponents of power-law and phantom-power-law expansion and other cosmological observables. We improve data reported earlier in \citep{arXiv:1008.2182}.
We find that although the forms of potential and the field solution are different for quintessential scalar field \citep{Gumjudpai:2012hs} \citep{arXiv:1008.2182} and tachyonic field, however  the equation of state are identical  for both quintessential scalar field and tachyonic field. This is to say that, for quintessence and tachyonic field, the equation of state does not depend on type of the scalar field but depends only on form of expansion function of the scale factor.  The present value of dark energy equation of state predicted by quintessential and tachyonic normal power-law cosmology models do not match both WMAP7 datasets. We conclude that the usual power-law cosmology model with either quintessence or with tachyonic field are excluded by these observational data. When considering the other case, the phantom power-law cosmology, the model predicts values of equation of state not much differ from observational results (for both quintessence and tachyonic cases), i.e. $w_{\phi, 0}  = -1.49^{+11.64}_{-4.08}$ (phantom power-law using WMAP7+BAO+$H_0$) and  $ w_{\phi, 0}  = -1.51^{+3.89}_{-6.72} $  (phantom power-law using WMAP7) compared to $w_{\phi, 0} = -1.34^{+1.74}_{-0.36}$ (WMAP7+BAO+$H_0$+SN):  and  $w_{\phi, 0} = -1.31^{+1.67}_{-0.38}$ (WMAP7+BAO+$H_0$+SN+time delay distance correction) \citep{arXiv:1001.4538}. From parametric plot in Fig. \ref{fig:1}, at $\b \lesssim -6$, $w_{\phi, 0}$ is in the expected range (-2, -1). We reconstruct the tachyonic potential in this scenario and we find that the dimensionless slope variable $\Gamma$ of our derived potential at present time is about 1.5. The phantom-power-law tachyonic potential found here reduced to $V= V_0\phi^{-2}$ in the limit $\Omega_{\m, 0} \rightarrow 0$.

%
\acknowledgments
We thank the referee for critical and useful comments. B.~G. is sponsored by a project number: BRG5380018 under the Basic Research Grant of the Thailand Research Fund (TRF). R. R. is funded via the TRF's Royal Golden Jubilee Doctoral Scholarship.

\section*{Appendix: Errors Analysis}
In calculating of the accumulated errors, we follow the procedure here. If $f$ is valued of answer in the form
\be
f = f(x_1, x_2, \ldots, x_n)
\ee
and $f_0$ is the value when $x_i$ is set to their measured values, then the value of $f_i$ is defined as
\be
f_i = f(x_1, \ldots, x_i + \sigma_i, \ldots, x_n)
\ee
This value of $f$ is the value with effect of error in variable $x_i$, that is $\sigma_i$. One can find square of the accumulated error from
\be
\sigma^2_{f}  =  \sum_i^n (f_i - f_0)^2
\ee
Hence giving the error of $f$ from accumulating effect from errors of $x_i$.


%

%

%

\begin{thebibliography}{99}


\bibitem[Abramo et. al. (2003)]{Abramo:2003cp}
Abramo, L.R.W., Finelli, F.:
  Phys. Lett. B \textbf{575}, (2003) 165

\bibitem[Aguirregabiria et. al. (2004)]{Aguirregabiria:2004xd}
Aguirregabiria, J.M., Lazkoz, R.:
  Phys. Rev. D \textbf{69}, (2004) 123502

\bibitem[Albrecht et. al. (1982)]{AlbrechtSteinhardt:1982a}
Albrecht, A., Steinhardt, P.J.: Phys. Rev. Lett. \textbf{48}, (1982) 1220

\bibitem[Allen (1999)]{Allen:1998vx}
Allen, R.E.: arXiv: astro-ph/9902042

\bibitem[Allen (2002)]{Allen2002}
Allen, S.W., Schmidt, R.W., Fabian, A.C.: Mon. Not. Roy. Astro. Soc. \textbf{334}, (2002) L11

\bibitem[Allen (2003)]{Allen:2003}
Allen, S.W., Schmidt, R.W., Fabian, A.C., Ebeling, H.: Mon. Not. Roy. Astro. Soc. \textbf{342}, (2003) 287

\bibitem[Allen et. al. (2004)]{Allen:2004cd}
Allen, S.W., {\it et al.}:
  Mon. Not. Roy. Astron. Soc.  \textbf{353}, (2004) 457

\bibitem[Amanullah (2010)]{Amanullah2010}
Amanullah, R., {\it et al.}:
  Astrophys. J. \textbf{716}, (2010) 712

\bibitem[Armendariz-Picon et. al. (2000)]{ArmendarizPicon:2000dh}
Armendariz-Picon, C., Mukhanov, V.F., Steinhardt, P.J.:
  Phys. Rev. Lett. \textbf{85}, (2000) 4438

\bibitem[Armendariz-Picon et. al. (2001)]{ArmendarizPicon:2000ah}
Armendariz-Picon, C., Mukhanov, V.F., Steinhardt, P.J.:
  Phys. Rev. D \textbf{63}, (2001) 103510

\bibitem[Astier (2006)]{Astier:2005qq}
Astier, P., {\it et al.} (SNLS Collaboration):
Astron. Astrophys. \textbf{447}, (2006) 31


\bibitem[Bagla et. al. (2003)]{Bagla:2002yn}
Bagla, J.S., Jassal, H.K., Padmanabhan, T.:
  Phys. Rev. D \textbf{67}, (2003) 063504

\bibitem[Bilicki et. al. (2012)]{Bilicki:2012ub}
Bilicki, M., Seikel, M.:
  Mon. Not. Roy. Astron. Soc. \textbf{425}, (2012) 1664


\bibitem[Caldwell (2002)]{Caldwell:1999ew}
Caldwell, R.R.:
  Phys. Lett. B \textbf{545}, (2002) 23

\bibitem[Caldwell et. al. (2003)]{Caldwell:2003vq}
Caldwell, R.R., Kamionkowski, M., Weinberg, N.N.:
  Phys. Rev. Lett.  \textbf{91}, (2003) 071301

\bibitem[Cataldo et. al. (2008)]{arXiv:0803.1086}
Cataldo, M., Mella, P., Minning, P., Saavedra, J.:
  Phys. Lett. B \textbf{662}, (2008) 314   

\bibitem[Coles et. al. (2002)]{coles}
Coles, P., Lucchin, F.: \textit{Cosmology, The Origin and Evolution of Cosmic Structure} (Wiley, 2nd Ed., 2002)

\bibitem[Copeland et. al. (2005)]{Copeland:2004hq}
Copeland, E.J., Garousi, M.R., Sami, M., Tsujikawa, S.:
  Phys. Rev. D \textbf{71}, (2005) 043003

\bibitem[Copeland et. al. (2006)]{Copeland:2006a}
Copeland, E.J., Sami, M., Tsujikawa, S.:
Int. J. Mod. Phys. D \textbf{15}, (2006) 1753


\bibitem[Dev et. al. (2001)]{Dev:2000du}
Dev, A., Sethi, M., Lohiya, D.:
Phys. Lett. B \textbf{504}, (2001) 207

\bibitem[Dev et. al. (2002)]{Dev:2002sz}
Dev, A., Safonova, M., Jain, D., Lohiya, D.:
Phys. Lett. B \textbf{548}, (2002) 12

\bibitem[Dev et. al. (2008)]{Dev:2008ey}
Dev, A., Jain, D., Lohiya, D.:
arXiv:0804.3491 [astro-ph]

\bibitem[Dolgov (1982)]{Dolgov1982}
Dolgov, A.D.: in \textit{The Very Early Universe}, ed. Gibbons, G., Hawking, S.W., Tiklos, S.T. (Cambridge University Press, 1982)

\bibitem[Dolgov (1997)]{Dolgov:1997a}
Dolgov, A.D.: Phys. Rev. D \textbf{55}, (1997) 5881



\bibitem[Ford (1987)]{Ford:1987a}
Ford,L.H.: Phys Rev D \textbf{35}, (1987) 2339

\bibitem[Fujii et. al. (1990)]{FujiiNishioka:1990}
Fujii, Y., Nishioka, T.: Phys. Rev. D \textbf{42}, (1990) 361


\bibitem[Garousi (2000)]{242}
Garousi, M.R.:
  Nucl. Phys. B \textbf{584}, (2000) 284

\bibitem[Garousi et. al. (2004)]{Garousi:2004uf}
Garousi, M.R., Sami, M., Tsujikawa, S.:
  Phys. Rev. D \textbf{70}, (2004) 043536

\bibitem[Goldhaber et. al. (2001)]{Goldhaber:2001a}
Goldhaber, G., {\it et al.} (The Supernova Cosmology Project Collaboration):
 Astrophys. J. \textbf{558}, (2001) 359

\bibitem[Gumjudpai et. al. (2012)]{GumjudpaiPowerlaw1}
Gumjudpai, B., Thepsuriya, K.:
  Astrophys. Space Sci. \textbf{342}, (2012) 537

\bibitem[Gumjudpai (2013)]{Gumjudpai:2012hs}
Gumjudpai, B.:
Mod. Phys. Lett. A. \textbf{28}, (2013) 1350122
 (arXiv:1307.4552)

\bibitem[Guth (1981)]{Guth:1981a}
Guth, A.H.: Phys. Rev. D \textbf{23}, (1981) 347




\bibitem[Jain et. al. (2003)]{Jain2003}
Jain, D., Dev, A., Alcaniz, J.S.: Class. Quan. Grav. \textbf{20}, (2003) 4163


\bibitem[Kaeonikhom et. al. (2011)]{arXiv:1008.2182}
Kaeonikhom, C., Gumjudpai, B., Saridakis, E.N.:
  Phys. Lett. B \textbf{695}, (2011) 45  

\bibitem[Kaplinghat et. al. (1999)]{Kaplinghat:1998wc}
Kaplinghat, M., Steigman, G., Tkachev, I., Walker, T.P.:
  Phys. Rev. D \textbf{59}, (1999) 043514

\bibitem[Kaplinghat et. al. (2000)]{Kaplinghat:2000zt}
Kaplinghat, M., Steigman, G., Walker, T.P.:
  Phys. Rev. D \textbf{61}, (2000) 103507

\bibitem[Kolb (1989)]{Kolb1989}
Kolb, E.W.: Astrophys. J. \textbf{344}, (1989) 543

\bibitem[Komatsu et. al. (2011)]{arXiv:1001.4538}
Komatsu, E., {\it et al.} (WMAP Collaboration):
  Astrophys. J. Suppl. \textbf{192}, (2011) 18  

\bibitem[Kumar (2012)]{Kumar2011}
Kumar, S.: Mon. Not. Roy. Astron. Soc. \textbf{422}, (2012) 2532

\bibitem[Kutasov (2003)]{Kutasov:2003er}
Kutasov, D., Niarchos, V.:
  Nucl. Phys. B \textbf{666}, (2003) 56


\bibitem[Larson et. al. (2011)]{Larson:2010gs}
Larson, D., {\it et al.}:
  Astrophys. J. Suppl.  \textbf{192}, (2011) 16

\bibitem[Linde (1982)]{Linde:1982a}
Linde, A.D.: Phys. Lett. B \textbf{108},(1982) 389

\bibitem[Lohiya et. al. (1996)]{Lohiya:1996wr}
Lohiya, D., Mahajan, S., Mukherjee, A., Batra, A.:
  arXiv:astro-ph/9606082

\bibitem[Lohiya et. al. (1999)]{Lohiya:1998tg}
Lohiya, D., Sethi, M.:
  Class. Quant. Grav. \textbf{16}, (1999) 1545

\bibitem[Lucchin et. al. (1985)]{Lucchin}
Lucchin, F., Matarrese, S.: Phys. Rev. D \textbf{32}, (1985) 1316


\bibitem[Manheim et. al. (1990)]{Manheim}
Manheim, P., Kazanas, D.: Gen. Rel. Grav. \textbf{22}, (1990) 289

\bibitem[Masi et. al. (2002)]{Masi:2002hp}
Masi, S., {\it et al.}:
Prog. Part. Nucl. Phys. \textbf{48}, (2002) 243

\bibitem[Melia et. al. (2012)]{Melia:2011fj}
Melia, F., Shevchuk, A.:
 Mon. Not. Roy. Astron. Soc.  \textbf{419}, (2012) 2579




\bibitem[Padmanabhan (2002)]{Padmanabhan:2002cp}
Padmanabhan, T.:
  Phys. Rev. D \textbf{66}, (2002) 021301

\bibitem[Padmanabhan (2005)]{Padmanabhan:2004av}
Padmanabhan, T.:
 Curr. Sci. \textbf{88}, (2005) 1057

\bibitem[Padmanabhan (2006)]{Padmanabhan:2006a}
Padmanabhan, T.:
 AIP Conf. Proc. \textbf{861}, (2006) 179

\bibitem[Peebles (1993)]{Peebles:1994xt}
Peebles, P.J.E.:
  \textit{Principles of Physical Cosmology},
  (Princeton University Press, 1993)

\bibitem[Peebles et. al. (2003)]{PeeblesRatra:2003}
Peebles, P.J., Ratra, B.: Rev. Mod. Phys. \textbf{75}, (2003) 559

\bibitem[Perlmutter et. al. (1998)]{Perlmutter:1997zf}
Perlmutter, S., {\it et al.}  (Supernova Cosmology Project Collaboration):
  Nature \textbf{391}, (1998)  51

\bibitem[Perlmutter et. al. (1999)]{Perlmutter:1999a}
Perlmutter, S., {\it et al.} (Supernova Cosmology Project Collaboration):
  Astrophys. J. \textbf{517}, (1999) 565



\bibitem[Rapetti et. al. (2005)]{Rapetti:2005a}
Rapetti, D., Allen, S.W., Weller, J.:
  Mon. Not. Roy. Astron. Soc.  \textbf{360}, (2005) 555

\bibitem[Rastkar et. al. (2012)]{SetareAstrophysicsSpace}
Rastkar, A.R., Setare, M.R., Darabi, F.: Astrophys. Space Sci. \textbf{337}, (2012) 487

\bibitem[Riess et. al. (1998)]{Riess:1998cb}
Riess, A.G., {\it et al.} (Supernova Search Team Collaboration):
 Astron. J. \textbf{116}, (1998) 1009

\bibitem[Riess (1999)]{Riess:1999ar}
Riess, A.G.:
 arXiv: astro-ph/9908237

\bibitem[Riess et. al. (2004)]{RiessGold2004}
Riess, A.G., {\it et al.} (Supernova Search Team Collaboration):
 Astrophys. J. \textbf{607}, (2004) 665

\bibitem[Riess et. al. (2007)]{Riess:2007a}
Riess, A.G., {\it et al.}:
 Astrophys. J. \textbf{659}, (2007) 98


\bibitem[Sahni et. al. (2000)]{SahniSta2000}
Sahni, V., Starobinsky, A.: Int. J. Mod. Phy. D \textbf{9}, (2000) 373

\bibitem[Sato (1981)]{Sato:1981a}
Sato, K.: Mon. Not. Roy. Astro. Soc. \textbf{195}, (1981) 467

\bibitem[Scranton et. al. (2003)]{Scranton:2003in}
Scranton, R., {\it et al.} (SDSS Collaboration):
  astro-ph/0307335

\bibitem[Sen (2002a)]{Sen:2002a}
Sen, A.:
  JHEP \textbf{0204}, (2002a) 048

\bibitem[Sen (2002b)]{Sen:2002b}
Sen, A.:
  JHEP \textbf{0207}, (2002b) 065



\bibitem[Starobinsky (1980)]{inflation}
Starobinsky, A.A.: Phys. Lett. B \textbf{91}, (1980) 99

\bibitem[Setare et. al. (2012)]{SetareDarabi:2012}
Setare, M.R., Darabi, F.: Gen. Rel. Grav. \textbf{44}, (2012) 2521

\bibitem[Sethi et. al. (2005)]{Sethi2005}
Sethi, G., Dev, A., Jain, D.: Phys. Lett. B \textbf{624}, (2005) 135

\bibitem[Sethi et. al. (1999)]{Sethi:1999sq}
Sethi, M., Batra, A., Lohiya, D.:
Phys. Rev. D \textbf{60}, (1999) 108301

\bibitem[Shapiro et. al. (2009)]{ShapiroSola:2009}
Shapiro, I.L., Sola, J.:
 Phys. Lett. B \textbf{682}, (2009) 105

\bibitem[Sola et. al. (2005)]{varcc}
Sola, J., Stefancic, H.:
 Phys. Lett. B \textbf{624}, (2005) 147


\bibitem[Tegmark et. al. (2004)]{Tegmark:2004a}
Tegmark, K., {\it et al.}  (SDSS Collaboration):
  Phys. Rev. D \textbf{69}, (2004) 103501

\bibitem[Tonry et. al. (2003)]{Tonry:2003a}
Tonry,J.L., {\it et al.} (Supernova Search Team Collaboration):
 Astrophys. J. \textbf{594}, (2003) 1






\bibitem[Wei (2001)]{astro-ph/0405368}
Wei, Y.-H.:
  astro-ph/0405368  

\bibitem[Weinberg (1989)]{cosmoconstant}
Weinberg, S.: Rev. Mod. Phys. \textbf{61}, (1989) 1




\bibitem[Zhu et. al. (2008)]{Zhu:2007tm}
Zhu, Z.H., Hu, M., Alcaniz, J.S., Liu, Y.X.:
Astron. and Astrophys. \textbf{483}, (2008) 15



\end{thebibliography}
%

\end{document}